\documentclass[pra,aps,floatfix,showpacs,tightenlines,superscriptaddress,amsmath,amssymb,showkeys,10pt,nofootinbib]{revtex4-2}
\usepackage{xcolor}
\usepackage{graphicx}

\usepackage{hyperref}

\newcommand{\be}{\begin{equation}}\newcommand{\ee}{\end{equation}}
\newcommand{\bea}{\begin{eqnarray}}\newcommand{\eea}{\end{eqnarray}}
\newcommand{\brr}{\begin{array}}\newcommand{\err}{\end{array}}
\newcommand{\bit}{\begin{itemize}}\newcommand{\eit}{\end{itemize}}
\newcommand{\ben}{\begin{enumerate}}\newcommand{\een}{\end{enumerate}}

\newcommand{\bbm}{\begin{bmatrix}}\newcommand{\ebm}{\end{bmatrix}}
\newcommand{\ba}{\begin{array}}
\newcommand{\ea}{\end{array}}

\newtheorem{mydef}{Definition}
\newtheorem{Lemma}{Lemma}
\newcommand{\bd}{\begin{mydef}} \newcommand{\ed}{\end{mydef}}
\newcommand{\bthe}{\begin{theorem}} \newcommand{\ethe}{\end{theorem}}
\newcommand{\ble}{\begin{Lemma}} \newcommand{\ele}{\end{Lemma}}

\newcommand{\dr}{\mathrm{d}}

\def\ha{\frac{1}{2}}

\def\lan{\langle}
\def\lf{\left}

\def\pa{\partial}\def\ran{\rangle}

\def\ri{\right}
\def\al{\alpha}

\def\la{\lambda}

\def\1{{_{1}}}\def\2{{_{2}}}

\def\noHe0{:\;\!\!\;\!\!:H_e(0):\;\!\!\;\!\!:}
\def\noHm0{:\;\!\!\;\!\!:H_\mu(0):\;\!\!\;\!\!:}

\def\lan{\langle}
\def\lf{\left}

\def\pa{\partial}\def\ran{\rangle}

\def\ri{\right}

\def\al{\alpha}

\def\la{\lambda}

\def\1{{_{1}}}\def\2{{_{2}}}

\begin{document}

\title{Aging properties of the voter model with long-range interactions}

\author{Federico Corberi}
\email{fcorberi@unisa.it}
\affiliation{Dipartimento di Fisica, Universit\`a di Salerno, Via Giovanni Paolo II 132, 84084 Fisciano (SA), Italy}
\affiliation{INFN Sezione di Napoli, Gruppo collegato di Salerno, Italy}

\author{Luca Smaldone}
\email{lsmaldone@unisa.it}
\affiliation{Dipartimento di Fisica, Universit\`a di Salerno, Via Giovanni Paolo II 132, 84084 Fisciano (SA), Italy}
\affiliation{INFN Sezione di Napoli, Gruppo collegato di Salerno, Italy}

\begin{abstract}
We investigate the aging properties of the one-dimensional voter model with long-range interactions in its ordering kinetics. In this system, an agent $S_i=\pm 1$ positioned at a lattice vertex $i$, copies the state of another one located at a distance $r$, selected randomly with a probability $P(r) \propto r^{-\alpha}$. Employing both analytical and numerical methods, we compute the two-time correlation function $G(r;t,s)$ ($t\ge s$) between the state of a variable $S_i$ at time $s$ and that of another one, at distance $r$, at time $t$. At time $t$, the memory of an agent of its former state at time $s$, expressed by the {\it autocorrelation function} $A(t,s)=G(r=0;t,s)$, decays algebraically for $\alpha >1$ as $[L(t)/L(s)]^{-\lambda}$, where $L$ is a time-increasing coherence length and $\lambda $ is the Fisher-Huse exponent. We 
find $\lambda =1$ for $\alpha >2$, and $\lambda =1/(\alpha-1)$ for $1<\alpha \le 2$. 
For $\alpha \le 1$, instead, there is an exponential decay, as in mean-field. Then, at variance with what is known for the related Ising model, here we find that $\lambda $ increases upon decreasing $\alpha$. The space-dependent correlation $G(r;t,s)$ obeys a scaling symmetry $G(r;t,s)=g[r/L(s);L(t)/L(s)]$ for $\alpha >2$. Similarly, for $1<\alpha \le 2$ one has $G(r;t,s)=g[r/{\cal L}(t);{\cal L}(t)/{\cal L}(s)] $, where now the length ${\cal L}$ regulating two-time correlations differs from the coherence length as ${\cal L}\propto L^\delta$, with $\delta=1+2(2-\alpha)$.

\end{abstract}

\maketitle
\section{Introduction}

When a system undergoes a phase-ordering process, such as a ferromagnet quenched below its critical temperature, the non-equilibrium character of the kinetics is clearly manifested by the aging properties displayed by two-time quantities. Typically, the
two-point/two-time correlation function, between two points at distance $r$, 
\begin{equation}
	G(r;t,s)=\langle S_i(t)S_j(s)\rangle,
\end{equation}
where $\{S_i\}$ is the collection of spin variables 
describing the elementary magnetic moments at lattice site $i$, and $t,s$ ($t>s$) are two generic times after
the quench, takes an additive form~\cite{Bray94}
\begin{equation}
	G(r;t,s)=G_{eq}(r;t-s)+G_{ag}(r;t,s).
	\label{split}
\end{equation} 
On the r.h.s., the former is a time-translation invariant contribution provided by fast degrees of freedom that locally equilibrate with the thermal bath. The latter term represents what is left over, namely the 
genuinely off-equilibrium effects. A physical intuition can be more easily achieved for scalar systems, i.e. $S_i=\pm 1$, to which we will always restrict in the following. In this case the post-quench evolution is
characterized by the formation and growth of ordered domains of characteristic size $L(t)$, whose interior is representative of the two, symmetry-related, pure states at low-temperature (say, with positive and negative magnetization), a phenomenon dubbed coarsening. Then, fast degrees involved in the first term on the r.h.s. of Eq.~(\ref{split}) are the so-called thermal islands, namely the ephemeral reversal of spins in small regions,
of the size of the equilibrium coherence length, caused by thermal fluctuations well inside the bulk of otherwise ordered domains. 
On the other hand $G_{ag}$ encodes the non-equilibrium 
motion of the domain walls. 

$G_{eq}$, being an equilibrium quantity, is generally quite well understood. 
In the following, therefore, we will be only interested
in $G_{ag}$, which can be conveniently isolated, for instance, by considering zero-temperature quenches,
because $G_{eq}\equiv0$ in this case, since thermal island cannot be activated with vanishing temperature. This is the situation we will consider in this paper and, for this reason, we will set $G_{ag}=G$ in the following, which
lightens the notation.  
This quantity
has been extensively studied~\cite{FurukawaJStatSocJpn,PhysRevB.40.2341,TJNewman_1990,Bray94,PhysRevE.52.6082,Corberi_2012,Chamon_2011,PhysRevE.65.046136,PhysRevLett.83.5054,PhysRevE.59.213} in systems with short-range interactions. Here it is generally found that $G$ obeys a scaling form
\begin{equation}
		G(r;t,s)=g\left (\frac{r}{L(s)};\frac{L(t)}{L(s)}\right ),
		\label{scalagt3}
	\end{equation} 
where $g$ is a scaling function.
Similarly to what is known in equilibrium critical phenomena, this structure reflects the physical fact that, at any time $t$, 
there is a unique physically relevant length $L$ and that the sole effect of the elapsing of time is the increase of 
such length. Then, the system is statistically similar to itself at any time, provided that $L$ is used as a unit to measure distances, as it can be seen by the first entry of $g$ in the above equation. The second entry expresses the fact that the passage of time, from $s$ to $t$, can be fully expressed by the fractional increase $L(t)/L(s)$ of the relevant length $L$.   
Eq.~(\ref{scalagt3}) implies
\begin{equation}
	A(t,s)=f\left (\frac{L(t)}{L(s)}\right ),
\end{equation}
for the so-called autocorrelation function $A(t,s)$, which is obtained 
letting $r=0$ in $G$, i.e. $A(t,s)=G(r=0;t,s)$, an important quantity frequently considered to study the aging properties~\cite{PhysRevE.102.020102,Corberi_2019,PhysRevE.93.052105,Corberi_2009,PhysRevE.74.041106}.
The scaling function has usually~\cite{Bray94,PhysRevB.38.373,PhysRevE.53.3073,PhysRevE.102.020102} the behavior
\begin{equation}
	f(x)\sim x^{-\lambda},\qquad \mbox{for }x\gg1,
	\label{defFH}
\end{equation}
where $\lambda$ is a non-trivial number also known as the Fisher-Huse exponent.

In recent years, there has been a growing interest in the investigation of long-range statistical models far from equilibrium, particularly concerning phase-ordering kinetics \cite{PhysRevE.49.R27,PhysRevE.50.1900,PhysRevE.99.011301,Corberi_2017,Corberi2019JSM,PhysRevE.103.012108,Corberi2021SCI,Corberi2023Chaos,Corberi2023PRE,PhysRevE.102.020102}, as well as other related aspects \cite{campa2009statistical,book_long_range,DauRufAriWilk}.
When long-range interactions are present, much less is known of the aging properties~\cite{PhysRevE.102.020102}.
This is because solvable models are scarce and numerical simulations difficult. For instance, considering phase-ordering, even in the simplest one-dimensional case, the Ising model, which was solved in the sixties~\cite{Glauber} for nearest-neighbors (NN)
interactions, cannot be exactly handled with space-dependent couplings. Numerical simulations, on the other hand, suffer of massive finite-size effects, caused by the extension of the interactions. 

In this respect, the possibility to study solvable models is mostly desired. A possibility is offered by the voter model, whose dynamical equations 
can be closed in any dimension~\cite{PhysRevA.45.1067,Frachebourg1996} and for every kind of
interaction~\cite{corberi2023kinetics,corsmal2023ordering}. Of course, despite being ferromagnetic in spirit, the model is not fully representative of 
physical magnets, because its kinetic rules generally lack detailed balance. However it reduces to the Ising model in space dimension $d=1$ with NN and is interested by a coarsening 
phenomenon also with extended interactions and
in $d=2$ as well. For $d\ge 3$, instead, there is no asymptotic coarsening.   

The voter model \cite{Clifford1973, Holley1975, liggett2004interacting, Theodore1986, Scheucher1988, PhysRevA.45.1067, Frachebourg1996, Ben1996, PhysRevLett.94.178701, PhysRevE.77.041121, Castellano09,krapivsky2010kinetic} was developed and popularized in the framework of ecological systems \cite{Clifford1973} as an invasion model, and later on in opinion dynamics \cite{Holley1975}, although some preliminaries were already investigated to study the spread of genetic correlations \cite{Kimura1964, 1970mathematics}. It can be described in this way: at each vertex of a lattice, an agent can adopt one of two options, typically represented as ``left" and ``right", which can equivalently be encoded in a binary variable $S_i=\pm1$, also called {\it spin} for the manifest similarity with the Ising model \cite{Zillio2005, Antal2006, Ghaffari2012, CARIDI2013216, Gastner_2018, Castellano09, Redner19}. In its original definition an agent's choice is influenced by a randomly selected neighboring agent, therefore with the probability of spin-flip being proportional to the fraction of opposite-spin neighbors.

Over the years, numerous modifications of the original model have been proposed to adapt it to explain various situations \cite{Mobilia2003,Vazquez_2004,MobiliaG2005,Dall'Asta_2007,Mobilia_2007,Stark2008,Castellano_2009,Moretti_2013,Caccioli_2013,HSU2014371,PhysRevE.97.012310,Gastner_2018,Baron2022}.
In such a context, a long-range variant of the voter model has been studied, where an agent  at a lattice vertex adopts the opinion of one located at a distance $r$ from him, selected randomly with a probability $P(r) \propto r^{-\alpha}$~ \cite{corberi2023kinetics,corsmal2023ordering}. In these papers the behavior of one-time quantities, specifically the equal time correlation function $G(r;t,t)$ was considered,
from which information on the coherence of the system and the approach to the ordered state can be extracted. 
Various regimes of $\alpha$ were analyzed, revealing a rich and diverse structure. For $d<3$ a critical value $\alpha_c(d)$ of $\alpha $ exists, such that for $\al > \al_c(d)$ the model behaves similarly to the NN case, with $L(t)$ growing as $\sqrt{t}$ until consensus, namely perfect ordering, is reached.
 It has been found that $\al_c(1)=3$ and $\al_c(2)=4$. Additionally, the correlation function exhibits dynamical scaling behavior, albeit with logarithmic corrections at large distances in the case $d=2$. Such corrections are due to the fact that $d=2$ is a critical dimensionality for the model. As $\alpha$ decreases below a certain threshold, there is the formation of partially ordered stationary states, which in the original NN model are only present for $d>2$. For $\al$ small enough, a behavior akin to mean-field ($\alpha=0$) is recovered.

In this paper we study this model in $d=1$, focusing on the aging properties that were not addressed before.
Although, clearly, the results derived here for the two-time
quantities discussed above are completely new and can by no means be inferred from the equal time quantities considered in~ \cite{corberi2023kinetics,corsmal2023ordering}, the present article must be regarded as a companion paper and many useful details can be found in the above references. Anyway, we will also provide the reader with the necessary context to ease 
the reading.  

In this paper, using both analytical and numerical methods to solve the exact closed dynamical equations (see, for $G$, Eq.~(\ref{eqc1ad1})), we calculate the two-time correlations and investigate their properties across various intervals of the parameter $\alpha$. A very brief and non-exhaustive
summary of our results follows:  
For $\alpha > 3$, the behavior is akin to one of the NN model, the autocorrelation function decays 
with $\lambda=1$ and $G$ obeys the scaling of Eq.~(\ref{scalagt3}), where $L(t)$ is a coherence
length that can be obtained from the equal-time
correlation function $G(r;t,t)$. 
In the range $2 < \alpha \leq 3$ and $1 < \alpha \leq 2$, $A(z)$ still displays a power-law decay characterized by $\lambda=1$ and $\lambda=(\alpha-1)^{-1}$, respectively.
Scaling of $G$ is still in the form~(\ref{scalagt3}), with the additional feature that,  only for $1<\alpha \le 2$, the scaling
length is not the coherence length.
When $\alpha \leq 1$, the autocorrelation function decays exponentially as in mean-field.

The paper is organized as follows: in Sec.~\ref{secmodel}, the model is presented. Secs.~\ref{agt3},\ref{ain23},\ref{ale2},\ref{ale1} are devoted to the analysis of the model in the regimes $\al>3$, $2< \al \leq 3$, $1 < \al \leq 2$, and $0\leq \al \leq 1$, respectively. In Sec.~\ref{secbounds} we briefly discuss how the found values of the exponent $\lambda$ fit into the known bounds for this quantity.
Finally, in \ref{secconcl}, a comparison with the Ising model is presented and our conclusions are drawn.
\section{The model} \label{secmodel}

In the long-range voter model considered here, a set of binary variables (spins), assuming the value $S_i=\pm 1$, are located on a $d$-dimensional lattice and interact with a probability $P(\ell)=\frac{1}{Z}\,\ell^{-\alpha}$, where $\ell$ is the distance between two of them. $Z$ is a normalization that, for $\alpha \le d$, depends on the number $N$ of spins. Focusing on the $1$-dimensional case -- the subject of this paper -- one has
\be
Z=2\sum_{\ell =1}^{N/2} \,\ell^{-\al}.
\label{eqZ} 
\ee
The factor $2$ appears because there
are two ways (on the right or on the left, say) to choose a spin at distance $\ell$ from a reference one. The sum extends up to $N/2$ because we consider a system of $N$ agents with periodic boundary conditions, where $N/2$ is the largest possible distance. From now on we will drop the summation range when 
sums over $\ell$ will be considered, and it will be understood that $\sum _\ell =\sum _{\ell=1}^{N/2}$. With this form of $P(\ell)$ it is clear that the moment of order $n$, $\lan \ell^n \ran \equiv \sum_\ell \ell^n P(\ell)$, diverges with $N$ for $\al\le n+1$. It can be expected that the moments with $n\le 2$, representing the normalization of probability, and a  characteristic length and length fluctuation related to $P(\ell)$, are relevant to the physical behavior. Therefore, their divergence in the thermodynamic limit, occurring at 
$\al =1,2,3$, respectively, will be
accompanied by a marked change of behavior of the model.
This is confirmed by the careful analysis of Ref. ~\cite{corberi2023kinetics} and by the results presented below. Such feature, therefore, is a direct consequence of the power-law behavior of $P(\ell)$.

The probability to flip a spin $S_i$ is
\be
w(S_i)=\frac{1}{2N} \, \sum _\ell P(\ell) \sum _{|k-i|=\ell}(1-S_iS_k) \, ,
\ee
where $k$ are the two sites at distance $\ell$ from $i$. Indeed, once the spin has been chosen (with probability $1/N$), the extraction of a spin $S_k$ at distance $\ell $ is an independent event occurring with probability $P(\ell)$, and we have to sum over all possible outcomes of $\ell$. Then, $S_i$
copies the state of $S_k$, which means that it changes state with probability
$\frac{1}{2}(1-S_iS_k)$.

Our aim is to find the two-time correlation functions  $G(r;t,s) =\langle S_i(t)S_{j}(s)\rangle$, where $r=|i-j|$ is the distance between the $i$-th and the $j$-th site and
$t$ and $s$ are two times with $t\ge s$. Following Ref.~\cite{Glauber} one has
$\frac{d}{dt}\langle S_{i_1}(t)S_{i_2}(t)\cdots S_{i_m}(t)S_{j_1}(s)S_{j_2}(s)\cdots S_{j_n}(s)\rangle=-2\langle S_{i_1}(t)S_{i_2}(t)\cdots S_{i_m}(t)S_{j_1}(s)S_{j_2}(s)\cdots S_{j_n}(s)\sum _{k=1}^mw(S_{i_k}(t))\rangle$
which, for $n=m=1$, provides
\begin{eqnarray}
\frac{d}{dt}\langle S_i(t)S_j(s)\rangle&=&-\langle S_i(t)S_j(s)\rangle+\sum _{\ell}P(\ell) \sum_{|k-i|=\ell}\langle S_k(t)S_j(s)\rangle \, ,
\label{eqc0}
\end{eqnarray}
 where time is measured in units of $N$ elementary moves, i.e. in Montecarlo steps.

Then, using the homogeneity of the model, Eq.~\eqref{eqc0}, can be written as
 \begin{eqnarray}
	\dot G (r;t,s) &=&- G (r;t,s) +\sum _\ell P(\ell)\left \{ G ([\![r-\ell]\!];t,s)
	+G ([\![r+\ell]\!];t,s)\right \}\, ,
	\label{eqc1ad1}
\end{eqnarray}
where the dot is a derivative with respect to $t$.
The double square-brackets indicate that we are using periodic boundary conditions. In other words, the topology of the system becomes that of a circle, which is obtained by gluing together the edges of the lattice. This means that
\be
[\![n]\!]=\left \{ \begin{array}{ll}
n & \mbox{if }|n|\le N/2 \\
N-|n| & \mbox{if } |n|>N/2
\end{array} \right.  \, .
\ee
Eq.~(\ref{eqc1ad1}) has to be solved with
the boundary condition $G(r;t,t)=C(r;t)$ where, proceeding
similarly~\cite{corberi2023kinetics}, one has that the equal time correlator $C$ obeys the evolution equation
 \begin{equation}
	\dot C (r,t) = \left \{ \begin{array}{lll}
	-2 C (r,t) +2\sum _\ell P(\ell) \left [C ([\![r-\ell]\!];t)+C([\![r+\ell]\!];t)\right ]&,&r>0,\\
	0&,& r=0,
	\end{array}
	\right .
	\label{eqc1}
\end{equation}
whose behavior has been studied in~\cite{corberi2023kinetics}. 
Let us remark that Eqs.~(\ref{eqc1ad1},\ref{eqc1}) are exact.
Notice that these two equations are identical apart for a trivial time rescaling (of a factor 2). However, the main difference is that Eq.~(\ref{eqc1}) does not apply at $r=0$,
because $C(r=0;t)\equiv 1$ does not evolve, whereas Eq.~(\ref{eqc1ad1}) holds down to $r=0$ as well. This fact, as we will see, makes a big difference. 

In the following, the correlation length
\begin{equation}
	L(t)=\frac{\sum _{r=1}^{N/2}  r\, C(r;t)}{\sum _{r=1}^{N/2} C(r;t)}
\end{equation}
will play an important role.

Before studying the aging properties of the model, it is useful to review the
general properties of the kinetics that were studied in~\cite{corberi2023kinetics}. Briefly, the behavior is different in the four sectors $\alpha>3$, $2<\alpha\le 3$, $1<\alpha \le 2$, and 
$0\le \alpha \le1$. For $\alpha >3$ 
a similar behavior as in the NN model is found, characterized by dynamical scaling and the {\it diffusive} character
$L(t)\propto t^{1/2}$. In the range
$2<\alpha \le 3$, $L(t)$ grows algebraically with an $\alpha$-dependent exponent and 
some scaling violations are observed.
For $1<\alpha \le 2$ coarsening is also observed, with an $\al$ dependent exponent of $L(t)$, but a non-trivial stationary state without consensus is reached, while for $0 \leq \al \leq 1$ no-macroscopic coarsening is present.

Before proceeding further, let us briefly discuss the idea behind selecting a power-law behavior for the probability $ P(\ell)$. Firstly, our focus primarily lies on the system's behavior for large $ t $. Because the domains' typical size increases over time, only the asymptotic behavior of $ P(\ell) $ for large $ \ell $ is relevant. Therefore, even more intricate forms of $ P(\ell)$, sharing the same asymptotic form $ \ell^{-\alpha}$, yield equivalent outcomes.
Forms of $ P(\ell) $ that asymptotically decay faster than $ \ell^{-\alpha}$ (e.g., exponentially) exhibit behavior akin to the case where $\alpha > 3$ in the present work, resembling that of the NN model. Conversely, for $P(\ell) $ decreasing slower than $ \ell^{-\alpha} $ (e.g., logarithmically), the system likely behaves as in the $ \alpha < 1 $ scenario of the present work, i.e. akin to the mean-field case.

\section{Case $\alpha >3$}
\label{agt3}

In this case we expect a behavior similar to the one observed
in the NN case, where dynamical scaling holds.
Therefore, for large $s$, we make the scaling ansatz of Eq.~(\ref{scalagt3}).
In order to lighten the notation, in the following we will set $L=L(t)$ and
$L_s=L(s)$. Plugging Eq.~(\ref{scalagt3}) in Eq.~(\ref{eqc1}),
letting $x=r/L_s$ and $z=L/L_s$, we get
\begin{equation}
\frac{\dot L}{L_s}\frac{\pa g(x,z)}{\pa z}=-g(x,z)+
	\sum_\ell P(\ell)\left \{g(|x-y|,z)+g([\![x+y]\!],z)\right \},
\end{equation}
where $y=\frac{\ell}{L_s}$. Upon McLaurin expanding the r.h.s. for small values of $y$ we obtain
\begin{equation}
z^{-1} \, \dot{L} \, L  \,\frac{\pa g(x,z)}{\pa z}=\langle \ell ^2\rangle \,\frac{\pa^2g(x,z)}{\pa x^2},
	\label{1dscal1}
\end{equation}
with $\langle \ell ^2\rangle=\sum _\ell \ell^2P(\ell)$.

Asking the explicit $t$-dependence to drop out of this equation we get
\begin{equation}
	L\ = \ D \, t^{1/2} \, ,
\end{equation}
so that Eq.~(\ref{1dscal1}) now reads
\begin{equation} \label{pde1}
z^{-1}\,\frac{\pa g(x,z)}{\pa z}=\frac{x_0^2}{2}\,\frac{\pa^2g(x,z)}{\pa x^2},
\end{equation}
with $x_0=\frac{2\langle \ell^2\rangle^{1/2}}{D}$.
The above equation should be solved with the
initial condition~\cite{corberi2023kinetics} at $t=s$
\begin{equation} \label{bcond}
 g(x,z=1)=\mbox{erfc}\left (\frac{x}{x_0}\right ) \, .
\end{equation}

\begin{figure}[htbp]
	\centering
	\includegraphics[width=1.\textwidth]{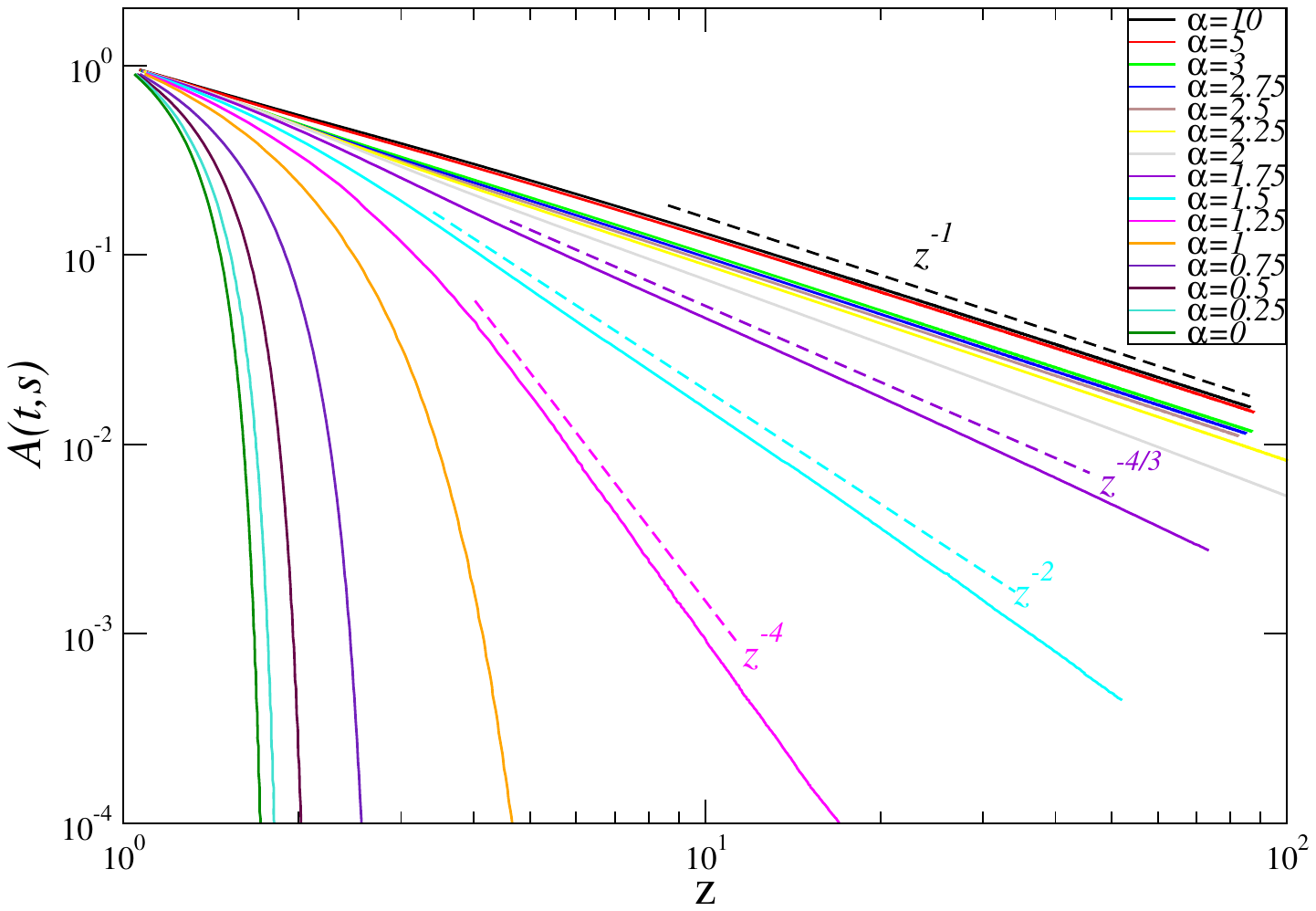}
	\caption{The autocorrelation $A(t,s)$ is plotted against $z=L/L_s$, for various values of $\alpha$, as detailed in the legend, for $s=1$. The dashed lines are the expected asymptotic (i.e. for large-$z$) behaviors~(\ref{lambda_gt_3},\ref{lambda_in23},\ref{lambda_in12}). 
	System size is $N=10^5$ for $\alpha\ge 2.5$, $N=3\cdot 10^5$ for $\alpha=2.25, 2$ and $\alpha\le 0.75$, and $N=10^6$ for the other values of $\alpha$.}
	\label{fig_A_vari_alfa}
\end{figure}

\begin{figure}[htbp]
	\centering
	\includegraphics[width=1.\textwidth]{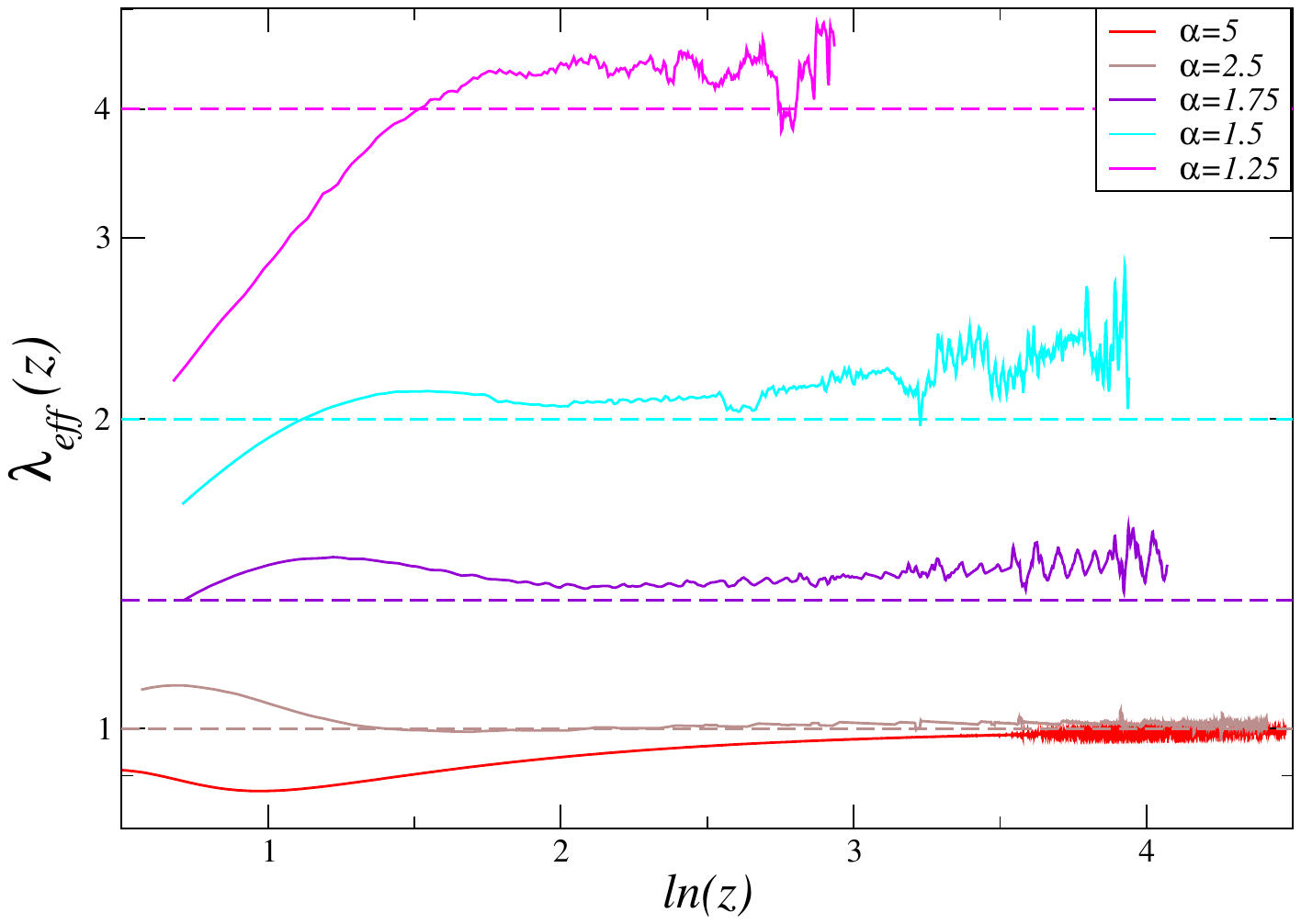}
	\caption{The effective exponent $\lambda_{eff}(z)$ is plotted against $z=L/L_s$, for various values of $\alpha$, as detailed in the legend, for $s=1$. This quantity has been obtained from the data of Fig.~\ref{fig_A_vari_alfa}. The dashed lines are the expected asymptotic (i.e. for large-$z$) values~(\ref{lambda_gt_3},\ref{lambda_in23},\ref{lambda_in12}).} 
	\label{fig_eff_exp}
\end{figure}

\begin{figure}[htbp]
	\centering
	\includegraphics[width=1.\textwidth]{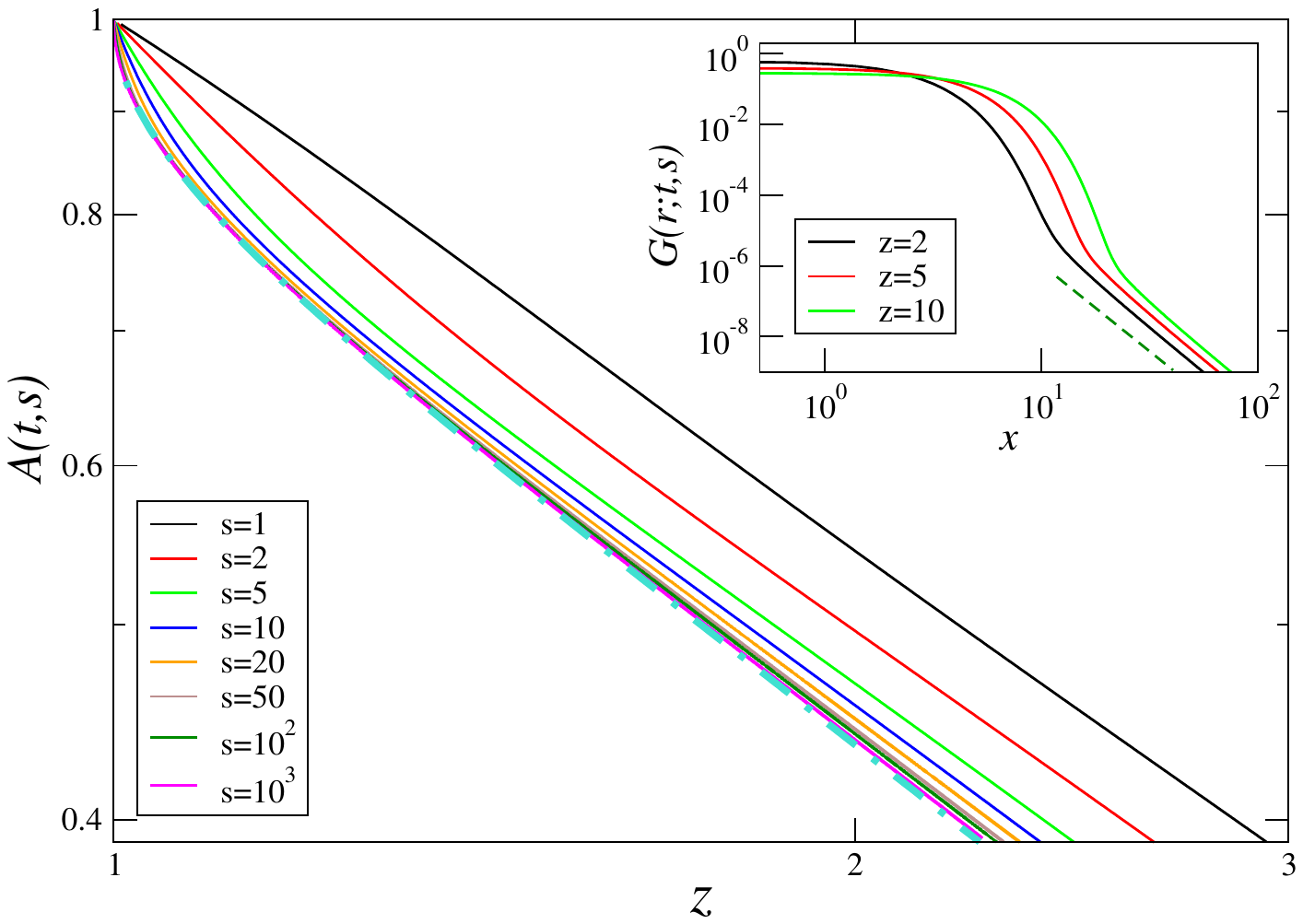}
	\caption{The autocorrelation function $A(t,s)$ is plotted against $z=L/L_s$ for $\alpha=5$ and different values of $s$, indicated in the legend.
	The system size is $N=10^3$. The dot-dashed bold turquoise line is the exact 
	result~(\ref{LZ}). 
	In the inset, $G(r;t,s)$ is plotted against $x=r/L_s$, still for $\alpha=5$, for $s=100$ and three values of $z$, see legend. The green dashed-line is the behavior $x^{-\alpha}$.}
	\label{fig_autocor_d1_alfa5_scal}
\end{figure}

In order to solve Eq.~(\ref{pde1}), we define $\zeta=z-1$ and we search a solution in the separated form
\be
g(\xi,\zeta) \ = \ f(\xi) \,h(\zeta) \, ,
\ee
where we introduced also the new variable $\xi \equiv x/x_0$.
The two functions satisfy the equations
\bea
\ha \, \frac{\dr^2 f(\xi)}{\dr \xi^2} & = & -\frac{\omega^2}{4} \, f(\xi) \, ,
\label{prima} \\[2mm]
\frac{1}{1+\zeta} \, \frac{\dr h(\zeta)}{\dr \zeta} & = & -\frac{\omega^2}{4} \, h(\zeta) \, . 
\label{seconda}
\eea
The constant, which physically must be negative, has been set to $-\frac{\omega^2}{4}$ for future convenience. The solutions of Eqs.~(\ref{prima},\ref{seconda}) are
\bea
f(\xi) & = & c_1(\omega) \cos \left(\frac{\omega  \xi }{\sqrt{2}}\right)+c_2(\omega) \sin \left(\frac{\omega  \xi }{\sqrt{2}}\right) \, , \\[2mm]
h(\zeta) & = & d(\omega) e^{-\frac{1}{4} \left(\frac{\zeta ^2}{2}+\zeta \right) \omega ^2} \, .
\eea
The general solution of Eq.~(\ref{pde1}) is then a linear combination 
\be
g(\xi,\zeta)  \ = \ \int^\infty_0 \!\! \dr \omega \, \lf[C_1(\omega) \cos \left(\frac{\omega  \xi }{\sqrt{2}}\right)+C_2(\omega) \sin \left(\frac{\omega  \xi }{\sqrt{2}}\right) \ri] \, e^{-\frac{1}{4} \left(\frac{\zeta ^2}{2}+\zeta \right) \omega ^2} \, , 
\ee
where $C_1(\omega)=c_1(\omega)d(\omega)$ and similarly for $C_2(\omega)$. The  initial condition \eqref{bcond} at $\zeta=0$ is obeyed if $C_2(\omega)=0$ and
\be
C_1(\omega) \ = \ \frac{4 F\left(\frac{\omega }{2 \sqrt{2}}\right)}{\pi ^{\frac{3}{2}} \,\omega } \, , 
\ee
where 
\be
F(x) \ = \ e^{-x^2} \, \int^x_0 \!\! \dr t \, e^{t^2} \, , 
\ee
is the \emph{Dawson function} \cite{olver2010nist}. With such choice, the autocorrelation function $A(z)$ reads
\be
A(z) \ = \ \frac{2}{\pi} \,  \arctan \left(\frac{\sqrt{2}}{\sqrt{z^2-1}}\right) \, , 
\label{LZ}
\ee
which is the same result as in the NN model \cite{Lippiello2000,Godrèche_2000}. Notice that for large $z$, i.e. $t \gg s$, this function goes as $1/z$, whereby, using Eq.~(\ref{defFH}),
one gets
\begin{equation}
	\lambda=1.
	\label{lambda_gt_3}
\end{equation}
This behavior can be clearly observed in Fig.~\ref{fig_A_vari_alfa}, where we plot the autocorrelation function 
against $z$ for several values of $\alpha$. $A(t,s)$ has been obtained by numerically integrating
Eq.~(\ref{eqc1ad1}) and singling out the $r=0$ value of $G(r;t,s)$. The result is,
in principle, exact, apart from the usual numerical errors (due to rounding and discretisation).
Clearly, in order to access
the asymptotic large-$z$ behavior, it 
is numerically convenient to use a sufficiently small value of $s$. This
is why we set $s=1$ in this figure.
Looking at the curves for $\alpha >3$
one can clearly see the 
$z^{-1}$ behavior, i.e. Eq.~(\ref{lambda_gt_3}). 
A more refined analysis is shown in Fig.~\ref{fig_eff_exp}, where the 
{\it effective exponent}
\begin{equation}
	\lambda_{eff}(z)=\frac{\dr \ln A(z)}{\dr \ln z}
\end{equation}
is plotted, for selected values of $\alpha$. The case with $\alpha=5$ shows that $\lambda _{eff}\simeq 1$
for $\ln z \gtrsim 4$, clearly confirming Eq.~(\ref{lambda_gt_3}).
Finally, in Fig.~\ref{fig_autocor_d1_alfa5_scal},
we verify that $A(t,s)$ is actually
a function of $z$ alone. In this case, by plotting curves for $A(t,s)$ at different $s$, we should see them
to superimpose when plotted against $z$.
This is indeed observed in the figure
for the largest values of $s$. Notice also that the curves approach the analytical solution~(\ref{LZ}), indicated by the dot-dashed bold turquoise line.
On the other hand, preasymptotic effects are important at least up to $s\simeq 20$, spoiling the collapse.
This implies that the data in Fig.~\ref{fig_A_vari_alfa}, obtained for $s=1$, are not in the scaling regime where Eq.~(\ref{scalagt3}) holds. As already explained, the choice $s=1$ is, however, the most convenient 
to access the large $z$ sector and measure $\lambda$, which is known~\cite{PhysRevE.53.3073} to be $s$-independent.

The general solution for the  correlation function of Eq.~(\ref{scalagt3}) is then obtained with 
\be
g(\xi,\zeta)  \ = \ \int^\infty_0 \!\! \dr \omega \,  \frac{4 F\left(\frac{\omega }{2 \sqrt{2}}\right)}{\pi ^{3/2} \omega } \,  \cos \left(\frac{\omega  \xi }{\sqrt{2}}\right) \, e^{-\frac{1}{4} \left(\frac{\zeta ^2}{2}+\zeta \right) \omega ^2} \, . 
\label{gexact}
\ee

The solution above is the asymptotic one for 
$s\to \infty$ (after taking $N\to \infty$).
It must be observed that, as discussed in~\cite{corberi2023kinetics}, important preasymptotic
corrections are expected at finite $s$ in the
large $x$ sector, for $x>x^*(s,z)$, with $x^*$
diverging both with $s\to \infty$ and $z\to \infty$. Repeating the argument in~\cite{corberi2023kinetics} it is possible to show that such corrections change the decay of 
$g(x,z)$ with $x$ from faster than exponential,
as it is in Eq.~(\ref{gexact}), to algebraic,
as $g(x,z)\propto x^{-\alpha }$. This can appreciated in the inset of Fig.~\ref{fig_autocor_d1_alfa5_scal}, where 
$G(r;s,t)$, obtained by the numerical solution of
Eq.~(\ref{eqc1ad1}), is plotted. Here $x^*$ is the
point (located around $x=10$ for $z=2$, for instance) where the downward bending curves 
turn into straight lines. It is seen that $x^*$
increases with $z$, and there is a similar trend with $s$ (not shown). Going asymptotic (in $s$ and/or in $z$) relegates the correction to farther and farther $x$-regions, where $G$ is smaller and smaller. Notice, indeed, that with the 
values of $s$ and $z$ chosen in the figure,
$x^*$ occurs when $G$ is already of order $10^{-6}$ or smaller.

\section{Case \boldmath{$2<\alpha \le 3$}}\label{ain23}

As noted before, the behavior found in the previous section  breaks down for $\alpha \le 3$, as $\langle \ell^2 \rangle$, appearing in Eq.~(\ref{1dscal1}), diverges. What happens here is that the McLaurin expansion used for $\alpha >3$ cannot be enforced, and a different analytical technique must then be used
to determine $G(r;t,s)$.

We now show that for $2< \alpha \le 3$ and large $s$, the correlation function $G(r;t,s)$ obeys a scaling form
\be
G(r;t,s) = A(z)g(x),\quad z\gg 1,
\label{factscal}
\ee
where $z$ have the same meaning as in Sec.~\ref{agt3} but now $x=r/L$ is defined differently. $g$ is a new scaling function, with
the large-$x$ behavior
\be
g(x)=(bx)^{-\alpha},
\label{unomenxbis}
\ee
where $b$ is a constant. 
Notice that Eq.~(\ref{factscal})
amounts to a factorization of the $z$ and $x$ dependencies, whereas for
$\alpha >3$ such property does not hold
(see Eq.~(\ref{scalagt3}) where $x$ and $z$ are mixed inside a unique scaling function). Let us also remark that such multiplicative form does not hold 
down to $z=1$, where different behaviors are found~\cite{corberi2023kinetics}.

\begin{figure}[htbp]
	\centering
	\includegraphics[width=1.\textwidth]{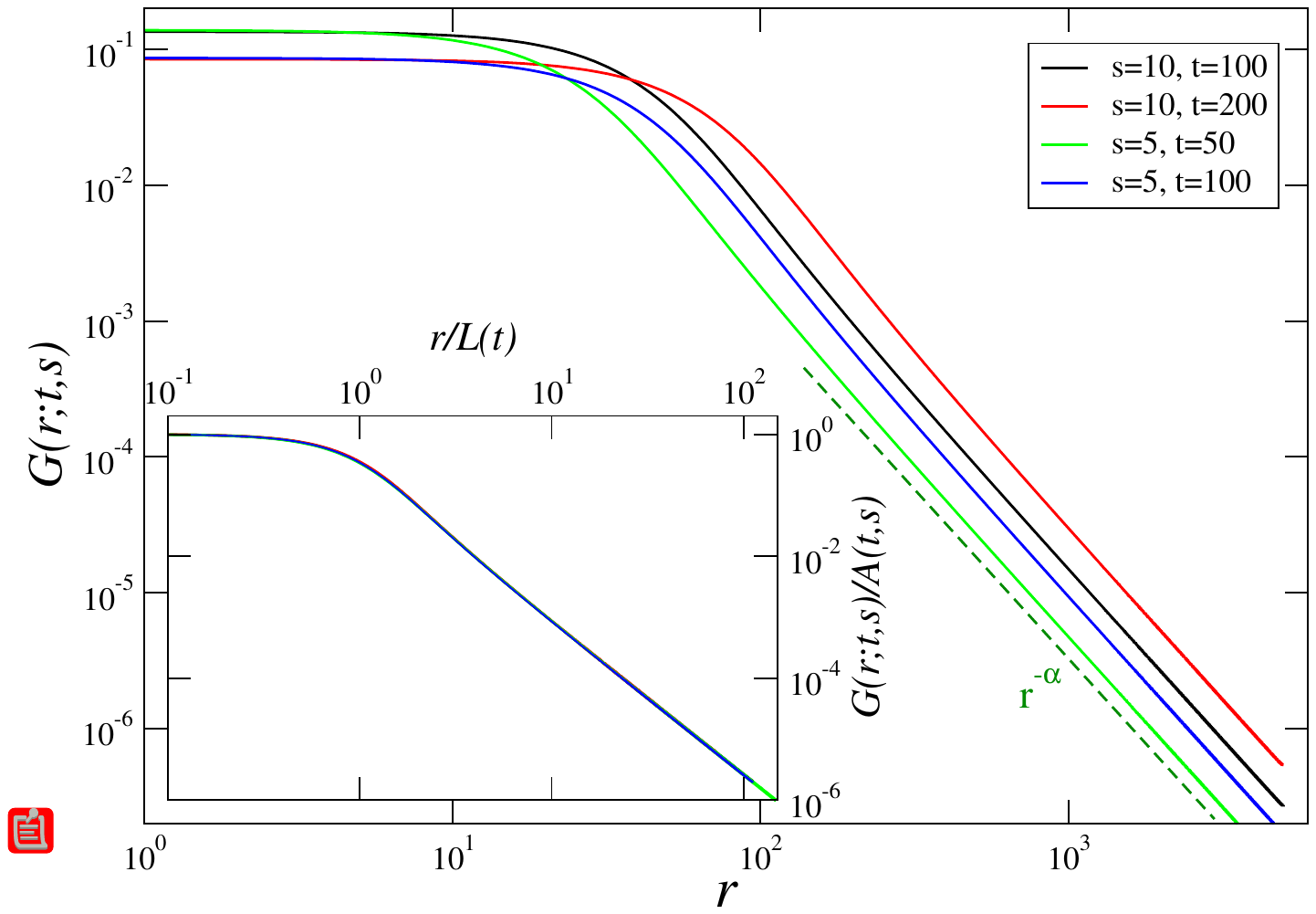}
	\caption{The correlation function
		$G(r;t,s)$ is plotted against $r$ on a double logarithmic plot, for $\alpha=2.5$ and different values of $s$ and $t$ (see legend). System size is $N=10^4$.
		The dashed line is the algebraic
		behavior of Eq.~(\ref{unomenxbis}).
		In the inset $A(z)^{-1}G(r;t,s)$
		is plotted against $x=r/L$.
		Curves for different $s$ are practically indistinguishable due to a remarkable collapse. 
	}
	\label{fig_twotimecor_d1_alfa2_5}
\end{figure}

Before showing the analytical consistency of the above forms~(\ref{factscal},\ref{unomenxbis}),
let us discuss the data presented in Fig.~\ref{fig_twotimecor_d1_alfa2_5}
where $G(r;t,s)$, obtained by numerically solving Eq.~(\ref{eqc1ad1}), is shown as a function of $r$, for $\alpha =2.5$ and different values of $t$ and $s$. In the main part of the figure it is clearly seen the algebraic decay~(\ref{unomenxbis}) at large $r$.
In the inset the curves are plotted
in a way that should produce collapse if the product form~(\ref{factscal})
holds, showing that this is indeed the case.

Returning now to the analytical calculations, let us 
insert the expression~(\ref{factscal}) into Eq.~(\ref{eqc1ad1}), thus having
\begin{equation}
	\frac{\dot L}{L}\,\left [g(x)zA'(z)-xg'(x)A(z)\right ]=S(x;t,z),
	\label{eqconS}
\end{equation}
where $S$ is the r.h.s. of Eq.~(\ref{eqc1ad1}).

In the range of $\alpha$ values considered here, the term containing
$C([\![r-\ell]\!],t)$ in Eq.~(\ref{eqc1ad1}) provides the dominant contribution to the r.h.s. for $r\gg 1$ when $s$ is large 
(it can be checked for consistency at the end of the calculation).
Writing the sum as an integral, such a term reads
\be
S(x;t,z)\simeq  2L(t)\int_{1/L(t)}^{N/2L(t)}\!\! \dr y\, P[yL(t)] G(|x-y|L(t);t,s),
\label{eqS}
\ee
with $y=\ell /L(t)$.

The integral is dominated by the region around the peak of the integrand at $y \approx x$, whose height is $P[xL(t)]G(0;t,s)=A(z)L(t)^{-\alpha}(2Z)^{-1}x^{-\alpha}$ and whose width, according to Eq.~(\ref{unomenxbis}), is of order $b^{-1}$. Hence
\be
S(x;t,z)\simeq   
\frac{L^{1-\alpha}x^{-\alpha}}{b Z}A(z)=b^{\alpha -1}Z^{-1}L^{1-\alpha}g(x)A(z).
\label{I0}
\ee

Then, plugging this expression into Eq.~(\ref{eqconS})  
we arrive at
\begin{equation}
	\frac{\dot L}{L}\left [g(x)zA'(z)-xg'(x)A(z)\right ] \ \simeq \ 
	b^{\alpha -1}Z^{-1}L^{1-\alpha}g(x)A(z),
	\label{newcons}
\end{equation}
where $A'(z)=\frac{dA(z)}{dz}$.
For large $r$, with the form~(\ref{unomenxbis}) it is $xg'(x)= -\alpha g(x)$, so that the 
$x$-dependence cancels out
indicating that the ansatz~(\ref{unomenxbis}) is consistent.
Imposing any explicit time-dependence to cancel out, the growth-law
\be
L(t)\ =\ B \, t^{1/(\alpha-1)}
\label{alphareg}
\ee
is obtained, where $B$ is a constant.
This is the same result previously found in~\cite{corberi2023kinetics}, where
$B$ was also estimated.
 
We remain with an equation for $A$
\begin{equation}
	z\,\frac{A'(z)}{A(z)}=-\la \, ,
	\label{zfrac} 
\end{equation}
where $\la $ is constant, that will turn out to be the Fisher-Huse exponent, so we use the same symbol.  The solution of Eq.~(\ref{zfrac}) is indeed
\be
A(z) \ = \ A_1 \, z^{-\la} \, ,\hspace{1.5cm}z\gg 1 \, , 
\label{eqA}
\ee
where $A_1$ is another constant.
The value of $\lambda$ is difficult to be computed analytically with the present technique, therefore we resort
to a numerical determination.
In Fig.~\ref{fig_A_vari_alfa} one can
see that the curves for $A(z)$ are practically superimposed for any value of $2< \alpha \le 3$. This possibly suggests that $A(z)$ is independent of
$\alpha$. If this is true, imposing continuity at $\alpha =3^+$ leads to  
\begin{equation}
	\la=1,
	\label{lambda_in23}
\end{equation}
which is indeed very well consistent with what one sees in Fig.~\ref{fig_A_vari_alfa}.
A further, more accurate check is provided in Fig.~\ref{fig_eff_exp}, where it is seen that, for $\alpha =2.5$, $\lambda_{eff}\simeq 1$ starting
as soon as from $\ln z\simeq 1.5$.

\section{Case \boldmath{$1 < \alpha \le 2$}}\label{ale2}

In this range of $\alpha$ the kinetics is characterized by a coarsening stage
ending up in a stationary state without consensus at a certain time $t_{stat}$
that diverges in the thermodynamic limit ~\cite{corberi2023kinetics}. 

\subsection{Coarsening stage}

We start by studying the aging properties in the coarsening stage. Let us remind that in Ref.~\cite{corberi2023kinetics} Eq.\eqref{eqc1} was explicitly solved, and the computation led to $L\propto t$.

We will show below that a multiplicative scaling similar to that of Eqs.~(\ref{factscal},\ref{unomenxbis})
is correct, however with a redefinition of the scaling variables
$x\to \tilde x=r/{\cal L}$ and $z\to \tilde z={\cal L}/{\cal L}_s$, where 
\begin{equation}
	{\cal L}=L^\delta
	\label{defbeta}
\end{equation}
 is an additional growing length regulating the spreading of 
two-time correlations, similarly to what $L$ does for the equal-time coherence. Then
\be
G(r;t,s) = {\cal A}(\tilde z)g(\tilde x),\quad \tilde z \gg 1,
\label{factscal1}
\ee
with $g$ of the form~(\ref{unomenxbis}).

 Proceeding as in Sec. \ref{ain23}, one arrives to an equation similar to Eq.~(\ref{eqconS}), which reads
\begin{equation} 
	\frac{\dot {\cal L}}{{\cal L}}\left [g(\tilde x) \tilde z {\cal A} '(\tilde z)-\tilde xg'(\tilde x){\cal A} (\tilde z)\right ]\ = S(\tilde x;t,\tilde z),
	\label{eqconS1}
\end{equation}
where again $S(\tilde x;t,\tilde z)$ is dominated
by the contribution due to the pronounced
peak of the argument of the sum around
$\ell \simeq r$ due to the maximum of 
$G([\![r-\ell]\!];t,s)$. 

Since this peak is narrow, one can take $P(\ell) \simeq P(r)$, so that 
\be
S(\tilde x;t,\tilde z) \ \simeq \  P(r) \, \sum_{r+\ell^*}^{r-\ell^*} \, G(|r-l|;t,s) \, , 
\label{eqSlsimr}
\ee
where $\ell^*$ is the typical length over which the peak decays moving away from the maximum at $\ell =r$. The area under the peak is proportional to 
the height $G(0;t,s)={\cal A}(\tilde z)$ of the
peak itself, times its width $\ell ^*$.
Note that $\ell^*$ may depend on time. Then
\be
S(\tilde x;t,\tilde z) \propto  r^{-\alpha}{\cal A}(\tilde z) \,\ell^*(t) \propto {\cal A}(\tilde z)\,\tilde x^{-\alpha}{\cal L}^{-\alpha}\ell^*(t)\propto \, {\cal A}(\tilde z)\,g(\tilde x)\,{\cal L}^{-\al} \, \ell^*(t)   \, , 
\label{eqSlsimr1}
\ee
the last passage holding for large $\tilde x$, where Eq.~(\ref{unomenxbis}) holds true.
Unfortunately, we have not been able to determine directly $\ell ^*$. However
we can infer its time-dependence by imposing that, once inserted the result~(\ref{eqSlsimr1}) into Eq.~(\ref{eqconS1}), the explicit time-dependence cancels out.
For an algebraic ${\cal L}$ this implies
${\cal L}^{-\alpha}\ell^*(t)\propto 1/t$. 
We thus arrive once more to an equation similar to Eq.\eqref{zfrac}, namely
\begin{equation}
	\tilde z \,\frac{{\cal A}'(\tilde z)}{{\cal A}(\tilde z)} \ = \ -\tilde \lambda \, ,
\end{equation}
with the solution
\begin{equation}
	{\cal A}(\tilde z) \ = \ A_2\,\tilde z^{-\tilde \lambda} \, ,
\end{equation}
where $A_2$ is a constant, implying
\begin{equation}
	A(z)\ = \ A_2\,z^{-\lambda} \, ,
\end{equation}
with $\lambda=\delta \tilde \lambda$. 
Still, $\la$ depends on unknown constants and it cannot be exactly computed, so we have to proceed numerically. In Fig.~\ref{fig_A_vari_alfa} it is seen that, in the range $1<\alpha \le 2$,
$A(z)$ decays algebraically as in Eq.~(\ref{defFH}) with an exponent $\lambda $ that keeps increasing 
as $\alpha $ is decreased. For $\alpha \le 1$, instead, the downward bending
of the curves signals an exponential decay of $A(z)$, as in mean-field. 
Then, the simplest interpolating form
between $\lambda=1$, for $\alpha=2$,
and $\lambda =\infty$ (i.e. exponential decay), for $\alpha=1$, is 
\begin{equation}
	\lambda =\frac{1}{\alpha -1} \, .
	\label{lambda_in12}
\end{equation}
Indeed, this expression is consistent with the data, as it can be seen in 
Fig.~\ref{fig_A_vari_alfa}.  Actually, a closer inspection shows that the slope of the curves is slightly larger than the one suggested by Eq.~(\ref{lambda_in12}). We have checked that this is due to important finite size effects. This can be discussed by looking at the effective exponent in Fig.~\ref{fig_eff_exp}.
In the case $\alpha=1.75$, for instance,
$\lambda_{eff}$ reaches a maximum at $\ln z\simeq 1.2$ and then decreases toward the value $4/3$ provided by Eq.~(\ref{lambda_in12}), indicated by the horizontal dashed line. However the descent stops at $\ln z =\ln z_{min}\simeq 2.1$, where $\lambda_{eff}\simeq 1.367$, and then the effective exponent starts increasing very slowly.
By repeating the calculation for different sizes we could verify that 
$z_{min}$ is delayed by increasing the system size $N$, and that $\lambda_{eff}(z_{min})$ approaches better the prediction of Eq.~(\ref{lambda_in12}). A similar behavior is observed for the other values of $\alpha \, \in \,  ]1,2[$. The curves presented in Figs.~\ref{fig_A_vari_alfa},~\ref{fig_eff_exp} correspond to the largest size
$N=10^6$ we were able to reach. In the other ranges of $\alpha $, namely for $\alpha >2$ or $\alpha <1$ finite size effects are much less severe (this is the reason for using such large sizes only for this intermediate range of $\alpha$). For $N=10^6$ we find 
$\lambda_{eff}(z_{min}) =1.367$ and $\lambda_{eff}(z_{min}) =2.055$ for $\alpha =1.75$ and $\alpha =1.5$, respectively. Using these values to estimate the Fisher-Huse exponent, indicating with $\lambda $ the expected value provided by Eq.~(\ref{lambda_in12}) there is a relative discrepancy $(\lambda_{eff}(z_{min})-\lambda)/\lambda=0.025$ or $0.027$ (for $\alpha=1.75$ and $\alpha =1.5$, respectively) which we consider well convincing. For $\alpha =1.25$ data for $\lambda_{eff}$ are too noisy to arrive to conclusions of comparable precision. Anyway, the discrepancy is reasonably less than $0.1$ also in this case, which is encouraging.

\begin{figure}[htbp]
	\centering
	\includegraphics[width=1.\textwidth]{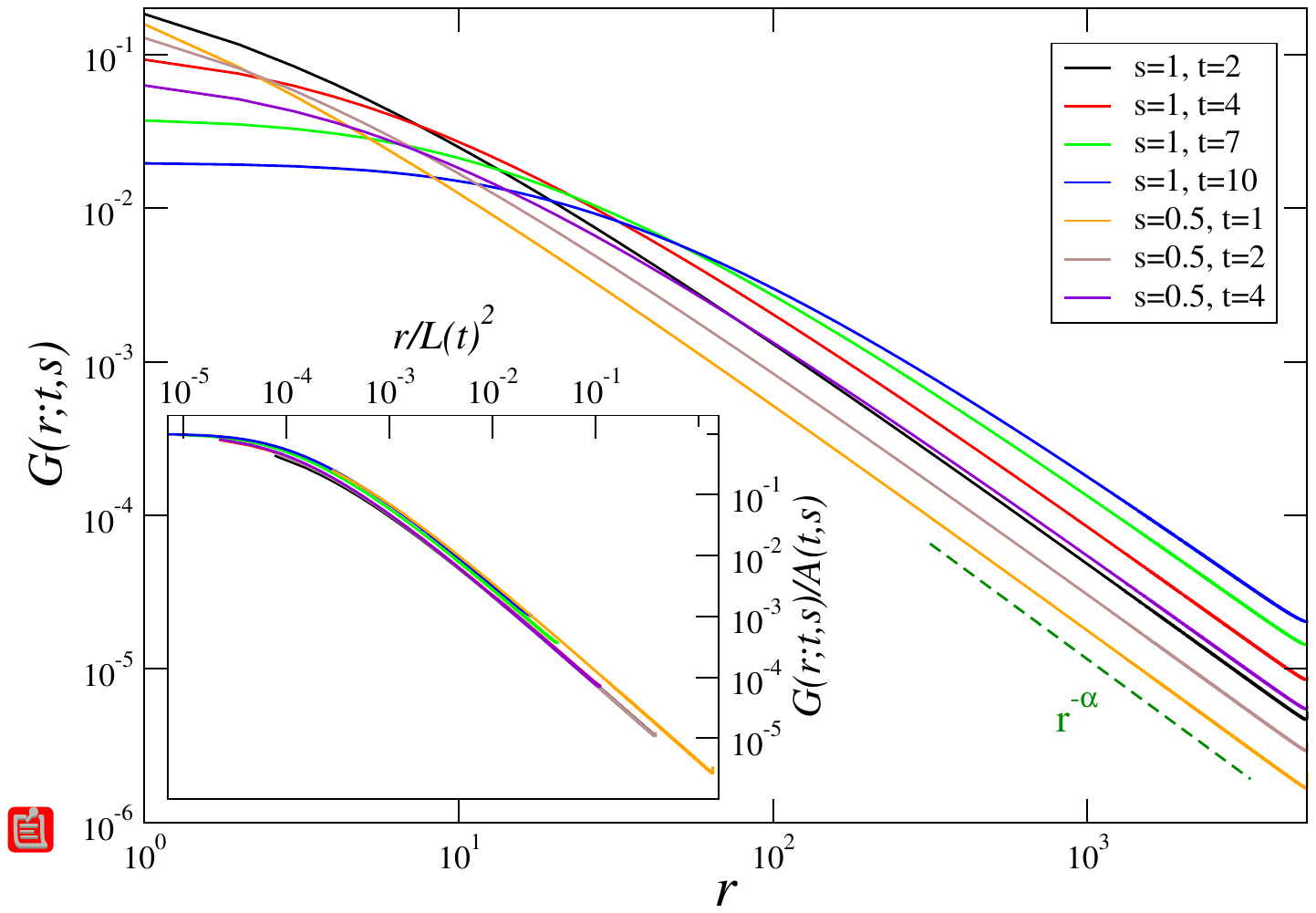}
	\caption{The correlation function
		$G(r;t,s)$ is plotted against $r$ on a double logarithmic plot, for $\alpha=1.5$ and different values of $s$ and $t$ (see legend). System size is $N=10^4$.
		The dashed line is the algebraic
		behavior of Eq.~(\ref{unomenxbis}).
		In the inset $A(z)^{-1}G(r;t,s)$
		is plotted against $x=r/L^\delta$, with $\delta=2$. 
	}
	\label{fig_twotimecor_d1_alfa1_5}
\end{figure}

The other exponent $\delta$, relating ${\cal L}$ to $L$ through Eq.~(\ref{defbeta}), can be determined 
numerically by searching data collapse
when plotting ${\cal A}^{-1}(\tilde z)G(r;t,s)$ against $\tilde x$, as implied by Eq.~(\ref{factscal1}), where 
the quantities involved are obtained by the numerical solution of Eq.~(\ref{eqc1ad1}). Similarly, one can
look for the analogous superposition upon plotting
$A(z)^{-1}G(r;t,s)$ against $\tilde x$, how it is done in Fig.~\ref{fig_twotimecor_d1_alfa1_5},
for $\alpha=1.5$. Here, in the inset,
such collapse is obtained with $\delta =2$. The superposition is good for sufficiently large values of $s$ and $t$,
with some preasymptotic correction which is visible, for instance, for $s=0.5$, $t=1$. In the main plot it can also be appreciated that the large-$r$
behavior~(\ref{unomenxbis}) is very neatly observed. Similar results are found for other values of $\alpha$ with
an exponent $\delta$ well consistent with 
\begin{equation}
	\delta=1+2(2-\alpha),
\end{equation}
in any case.

\subsection{Stationary states} \label{sec_stat}

We start noticing that, as already observed, Eqs.~(\ref{eqc1}) and (\ref{eqc1ad1}) are equal, apart from a trivial factor $2$ and the behavior in $r=0$. This means that
the form
\begin{equation}
	C_{stat}(r)\propto r^{-(2-\alpha)}
	\label{Cstat}
\end{equation} 
found in~\cite{corberi2023kinetics} for the 
one-time correlator at stationarity, would
produce $\dot G=0$ if inserted in Eq.~(\ref{eqc1ad1}). However this would not be true at $r=0$, due to the different form of the
equation. Indeed, indicating with $\tau=t-s$ the time difference, by definition it must be $G_{stat}(r=0;\tau)=A_{stat}(\tau)$, $A_{stat}$ being the autocorrelation in the stationary state, which is a decreasing function of $\tau$, whereas $C_{stat}$ is obviously time-independent. Hence the simplest 
ansatz one can do for $G_{stat}$ is
\begin{equation}
	G_{stat}(r;\tau)\simeq \left \{ \begin{array}{ll}
		A_{stat}(\tau), & \hspace{2cm} \mbox{for }r\ll r^* \\
		C_{stat}(r), & \hspace{2cm} \mbox{for }r\gg r^*,		\end{array}
	\right .
	\label{Gstat}
\end{equation}
with
\be
A_{stat}(\tau) \ \propto \  \tau^{-\la} \, .
\ee
$r^*$ can be found by matching
the two forms in the equation above, leading to
\begin{equation}
	r^*(t)\propto \tau^{\frac{\lambda}{2-\alpha}}.
	\label{rstar}
\end{equation}

We show below that, with the form~(\ref{Gstat}), once inserted in Eq.~(\ref{eqc1ad1}), one gets
\begin{equation}
	\dot G_{stat}(r;\tau)\simeq \left \{ \begin{array}{ll}
		\dot A_{stat}(\tau), & \hspace{2cm} \mbox{for }r\ll r^* \\
		0, & \hspace{2cm} \mbox{for }r\gg r^*,		\end{array}
	\right .
	\label{Gdotstat}
\end{equation}
for large $\tau$,
where now $\dot G_{stat}$ means $\dr G_{stat}/\dr \tau$, an similarly for $\dot A_{stat}$. This provides the consistency of our assumption~(\ref{Gstat}). Let us derive now
Eq.~(\ref{Gdotstat}), splitting the problem 
in the the two sectors. For $r\ll r^*$, the first 
term on the r.h.s. of Eq.~(\ref{eqc1ad1}) is
$G_{stat}(r;\tau)=A_{stat}(\tau)$ and the same value is taken by $G_{stat}([\![r-\ell]\!])$ and $G_{stat}([\![r+\ell]\!])$, for all $\ell < r^*-r$. Because of the normalization of $P$ (see Eq.~(\ref{eqZ})), the sum for $\ell < r^*-r$ cancels with the first piece on the r.h.s. of Eq.~(\ref{eqc1ad1}). Then, the value of the whole r.h.s. reads 
\begin{equation}
	A(\tau)\sum _{\ell=r^*-r}^{N/2}P(\ell)\simeq \frac{A(\tau)}{Z(\alpha -1)}\left [r^*(\tau)-r\right ]^{1-\alpha} \simeq \frac{A(\tau)}{Z(\alpha -1)}{r^*(\tau)}^{1-\alpha}.
\end{equation}
Using Eq.~(\ref{rstar}) to match the first line of Eq.~(\ref{Gdotstat}) provides the value
\begin{equation}
	\lambda=\frac{2-\al}{\al-1} \, .
	\label{lambdastat}
\end{equation} 
Now we turn to the sector $r\gg r^*$
In this case the first term on the r.h.s. of Eq.~(\ref{eqc1ad1}) is
$G_{stat}(r;\tau)\simeq C_{stat}(r>r^*)\simeq 0$
since $r^*$ is large. Similarly, $G_{stat}([\![r+\ell]\!])$
is even smaller. The term $G_{stat}([\![r+\ell]\!])$, instead, provides an appreciable contribution
$G_{stat}([\![r+\ell]\!])\simeq A_{stat}(\tau)$ to the sum in the region $r-r^*\le \ell \le r+r^*$. Since $r\gg r^*$ this contribution is approximately 
$2r^*P(r)A_{stat}(\tau)<2Z^{-1}{r^*}^{1-\alpha}A_{stat}(\tau)$,
which is small for $\alpha >1$. In conclusion, our assumption~(\ref{Gstat}), together with
the value~(\ref{lambdastat}), is consistent with the evolution equation~(\ref{eqc1ad1}).

\begin{figure}[htbp]
	\centering
	\includegraphics[width=1.\textwidth]{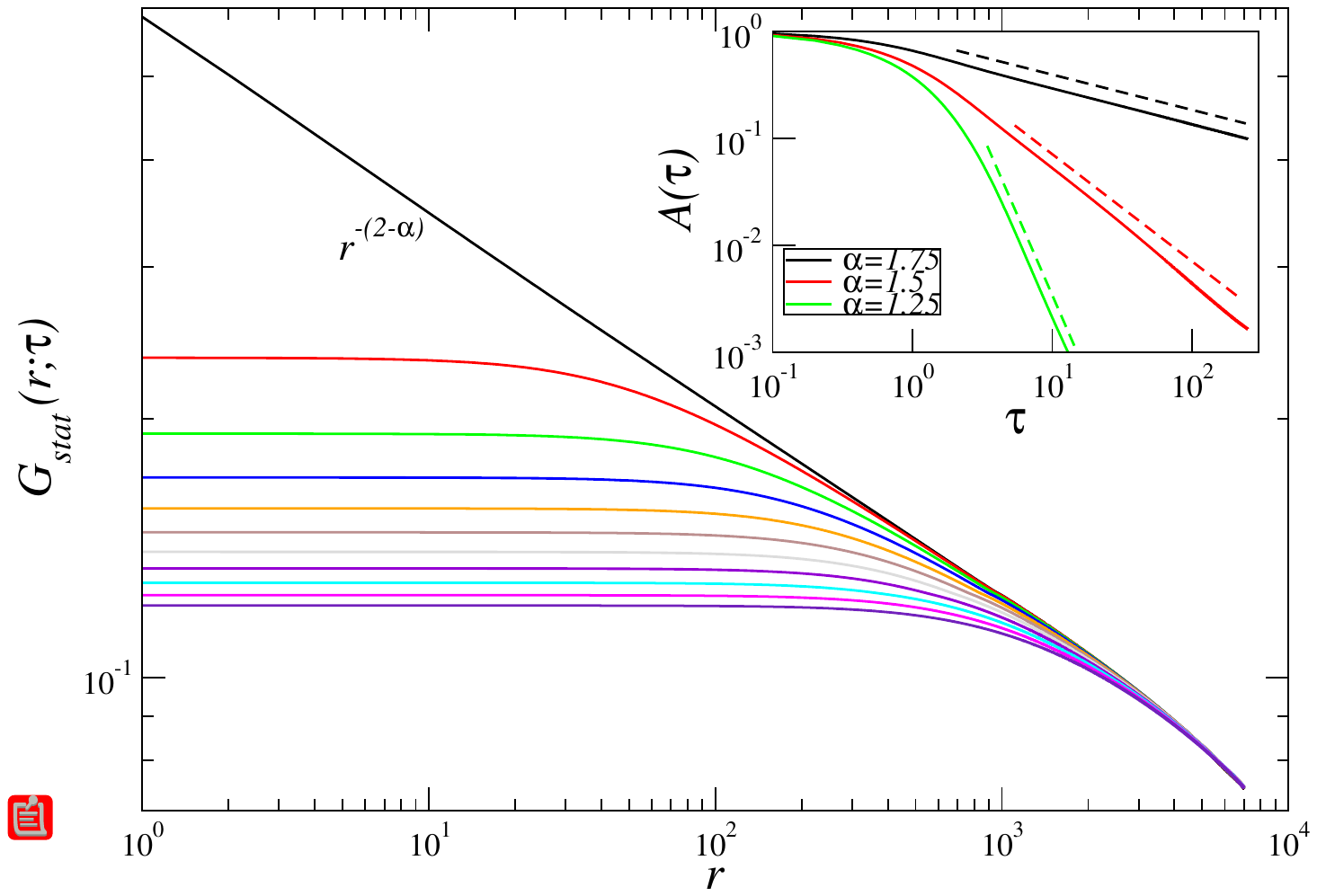}
	\caption{The stationary correlation function
		$G_{stat}(r;\tau)$ is plotted against $r$ on a double logarithmic plot, for $\alpha=1.75$ and different values of $\tau$, $\tau=0,25,50,\dots,500$, from top to bottom. In the inset, the autocorrelation function $A_{stat}(\tau)$ is plotted against $\tau$ in a double-logarithmic scale. Different curves correspond to different values of $\alpha$, see legend. System size is $N=10^5$.
		The dashed line is the algebraic
		behavior with $\lambda$ given in Eq.~(\ref{lambdastat}). 
	}
	\label{fig_stat}
\end{figure}

All the results derived above for the stationary
state are in agreement with the numerical solution
of Eq.~(\ref{eqc1ad1}), as it can be seen in Fig.~\ref{fig_stat}. Here, in the main part, 
the behavior~(\ref{Gstat}) can be appreciated. 
The top black curve is $C_{stat}(r)$, which decays as in Eq.~(\ref{Cstat}). Similar results are obtained for other values of 
$\alpha$ in $]1,2]$. In the inset, 
$A_{stat}(\tau)$ is plotted, which is very well consistent with the determination~(\ref{lambdastat}) of $\lambda$.

\section{Case \boldmath{$0 \le \alpha \le 1$}}\label{ale1}
For $\al \le 1$ there is not a real coarsening
stage~\cite{corberi2023kinetics}. Actually $L(t)$ initially increases, but the time domain of such
growth is microscopic and cannot by any mean be
extended, neither in the thermodynamic limit, because it is independent of $N$.
In a short time the system enters the stationary state and then, if finite, it goes to consensus.
Full ordering is only achieved in the thermodynamic limit. 

In this range of $\alpha $ we expect the system to share many features of the mean-field case, 
i.e. $\alpha =0$. In such limit, the master equation for the probability of a spin to be reversed $n$ times reads~\cite{Ben1996}
\begin{equation}
	\dot P_n(\tau)=\frac{1}{2}\left [P_{n-1}(\tau)-P_n(\tau)\right ],
\end{equation}
where $\dot P=\dr P/\dr \tau$, whose
solution is the Poissonian
\begin{equation}
	P_n(\tau)=\frac{\left (\frac{\tau}{2}\right )^n }{n!}\,e^{-\frac{\tau}{2}}.
\end{equation}
Starting from $P_n$, the autocorrelation function can be computed as
\begin{equation}
	A(\tau)=\sum_{n=0}^{N\tau}(-1)^nP_n(\tau)=e^{-\tau}.
	\label{Aisexp}
\end{equation}
Notice that the sum is bounded above by the maximum number of possible spin flips in the considered time interval. For large $N$ we have set this number equal to infinity to get the final result. 

\begin{figure}[htbp]
	\centering
	\includegraphics[width=1.\textwidth]{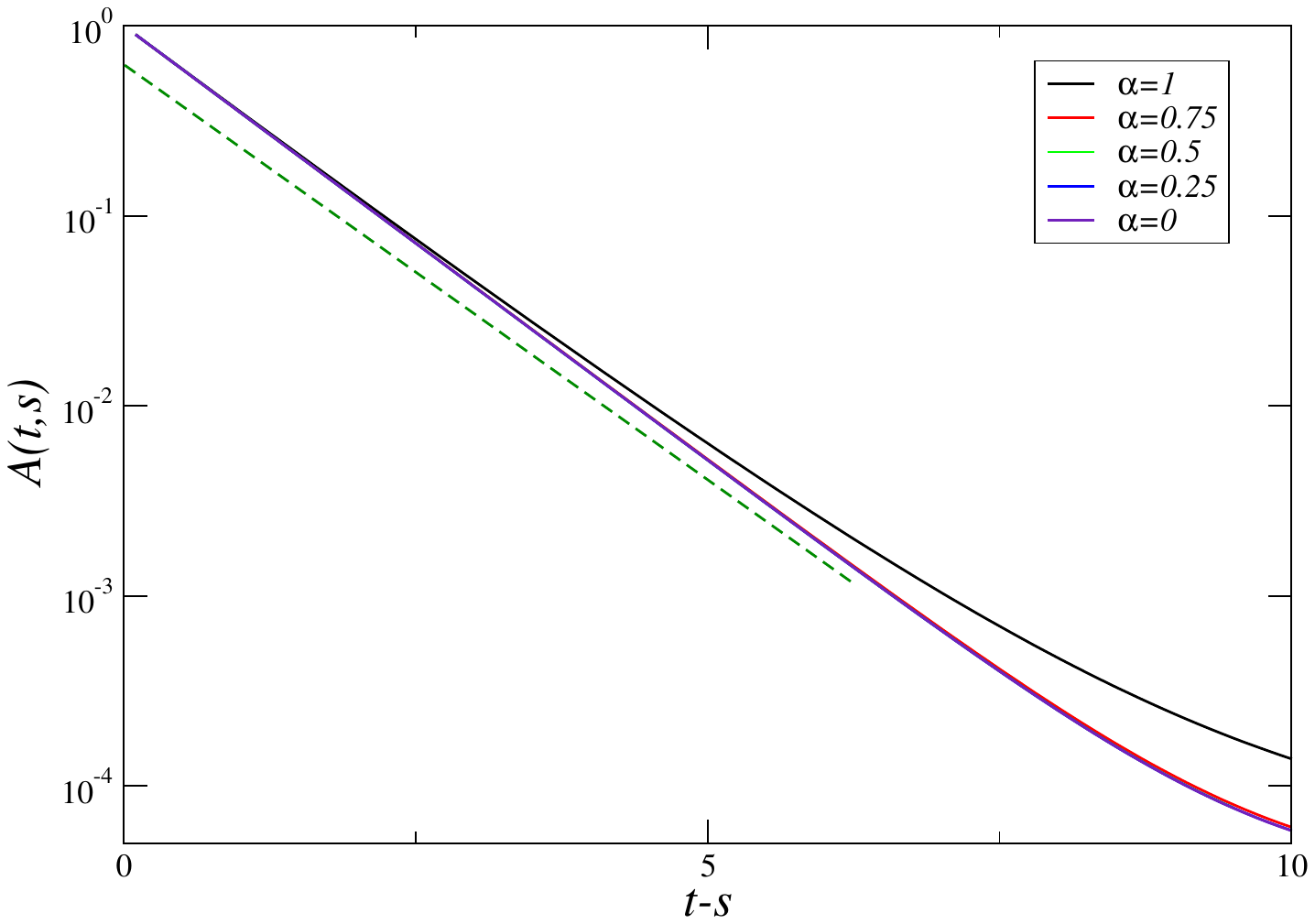}
	\caption{The autocorrelation function
		$A(\tau)$ is plotted against $\tau$ on a double logarithmic plot, for different values of $\al$, see legend. System size is $N=10^5$.
		The dashed line is the exponential
		behavior given in Eq.~(\ref{Aisexp}), shifted downwards of a small amount to make it visible. 
	}
	\label{fig_alfa_less_1_is_exp}
\end{figure}

In Fig.~\ref{fig_alfa_less_1_is_exp} we plot 
$A(t,s)$, as obtained by the numerical integration of Eq.~(\ref{eqc1ad1}). We see that
curves for any $\alpha<1$ collapse onto each other, including the mean-field case $\alpha=0$.
The behavior of the curves is very close by 
the one in Eq.~(\ref{Aisexp}), as it should, down to values of $A$ of order $10^{-3}$. At late times
a finite size effect is observed, in the form of 
a flattening of the curves. For $\alpha=1$ there is probably a logarithmic correction due to marginality. In conclusion, the mean-field behavior remains unchanged up to $\alpha =1^-$.

\section{\boldmath{$\lambda$} bounds.} \label{secbounds}

In~\cite{PhysRevB.38.373,PhysRevE.53.3073,PhysRevE.102.020102} the following bound
\begin{equation}
	\frac{d+\beta}{2}\le \lambda \le d
	\label{FHbound}
\end{equation}
was derived for the exponent $\lambda$, for a system undergoing phase-ordering in $d$ dimensions. Here $\beta $ is
an exponent regulating the small-$q$ behavior 
$S(q;t)\sim q^\beta$ of the structure factor $S$, the Fourier
transform of $C(r;t)$. 
The derivation of the lower bound is rather general
and only relies on the hypotheses that $C(r;t)$ obeys scaling. 
This is true in the present case except
in the range $2< \alpha \le 3$, where violations are observed~\cite{corberi2023kinetics}.
In the model considered insofar 
one has a large-$r$ fast decay~\cite{corberi2023kinetics} of $C$ for $\alpha >3$, turning to $C\sim r^{-\alpha}$ for $\alpha \le 3$, leading to $\beta=0$ for $\alpha >3$ and $\beta=\alpha -d$ for $\alpha \le 3$. The upper bound in Eq.~(\ref{FHbound}), instead, can only be derived 
assuming scaling with $\beta =0$, although in a non-rigorous way~\cite{PhysRevE.102.020102}.
Applying the above considerations to the model here discussed, we have
\begin{equation}
	\left \{ \begin{array}{ll}
	\frac{1}{2}\le \lambda \le 1\,, \quad & \mbox{for }\alpha >3 \\
	   \\
	\lambda \ge \frac{\alpha}{2}\,, \quad & \mbox{for }1<\alpha \le 2 \, .
	\end{array}
	\right . 
	\label{bound}
\end{equation}
For $\alpha >3$ we got $\lambda =1$ (Eq.~(\ref{lambda_gt_3})), and the first
line of Eq.~(\ref{bound}) is verified, saturating the upper bound. For $1<\alpha \le 2$, with the result~(\ref{lambda_in12}) the bound is also verified, and it is saturated right at $\alpha =2$. Let us recall that the bounds~(\ref{FHbound}) apply to a coarsening
system and, therefore, they do not apply for the
stationary states considered in Sec.~\ref{sec_stat}.

\section{Discussion and conclusions} \label{secconcl}

In this Article we have studied the aging properties of 
the one-dimensional voter model where an agent take the state of another at distance $r$, chosen with probability $P(r)\propto r^{-\alpha}$  . Besides its own interest, this model can be used as a proxy to the $1d$ Ising model with a coupling constant among spins $J(r)=P(r)$, quenched to zero temperature.
The difference is that spin $S_i$ in the Ising model take the sign of the Weiss field $H_i=\sum _{r}J(r)S_{i+r}$, which is 
an average over all the agents, weighted by $P(r)$, whereas
in the voter one single variable $S_{i+r}$ influences $S_i$. 
In a sense, the Ising model is a smooth version of the voter one.
Such averaging procedure has drastic consequences, changing profoundly the aging properties, as we explain below. In both models the value $\alpha =2$ separates a region $\alpha >2$ where the autocorrelation exponent conserves the NN value~\cite{Glauber} $\lambda =1$
from the region $\alpha \le 2$ where a different value, caused by
the long-range character, is observed. However this small-$\alpha $ value is 
$\lambda =1/2$ for Ising, whereas it is $\lambda =1/(\alpha -1)$
for voter (for $\alpha \le 1$ there is an exponential decay).
Then, there is an important qualitative difference between the 
two models: extending the range of interactions (i.e. lowering 
$\alpha$) increases the persistence of the memory of the Ising spins (i.e. $\lambda$ decreases), whereas the opposite occurs for voters. This suggests that a very different physical ordering mechanism is at work in the two systems. Indeed, in the Ising model, while for $\alpha >2$ the domain walls move as unbiased random walkers, for $\alpha \le 2$ there is a deterministic drift towards the closure of smaller domains~\cite{PhysRevE.102.020102}.
Then, the fraction of spins in large domains is extremely persistent and this increases the duration of memory. In the voter model there is, instead, a contrasting mechanism for $\alpha \le 2$. Namely, the possibility to interact with far-away spins, much beyond the size of domains, causes the reversal of
spins even in the bulk of correlated regions, producing the loss of memory. The fact that $\alpha =2$ is the limiting value below which this phenomenon becomes relevant is not by chance. Indeed, this is also the value of $\alpha $ below which in the thermodynamic limit the system does not reach consensus but 
gets trapped in a partially ordered stationary state. The reason for the formation of such stationary states is the same, a large amount of spin flips caused by long-distance interactions.

On the other hand, finding the same $\lambda $ exponent for $\alpha >2$ might suggest that the aging properties of the two models are similar. Then the rather precise results obtained
here for the voter model could shed some light on the properties
of the long-range Ising model, for which a closed equation akin to Eq.~(\ref{eqc1ad1}) cannot be derived. For example, 
we found that $A(z)$ has a unique expression (Eq.~(\ref{LZ})) 
for any $\alpha >3$. Trivially, the Ising model with NN shares 
the same expression, because with NN the two models can be mapped onto each other, but is it true that the same form is obtained for the Ising system also for sufficiently large values of $\alpha$? And, if yes,
does it holds true down to $\alpha =3$, as in the voter case?
Similar properties are obeyed also for the two-point correlation $G(r;t,s)$? 

This paper is the first where the aging properties of the long-ranged voter model have been considered. Even if we tried to provide a rather comprehensive panorama of the behavior of 
the two-time correlations in the whole range of $\alpha $ values,
a lot of questions remain unanswered. Firstly, since
a closed equation as Eq.~(\ref{eqc1ad1}) can be written in any 
spatial dimension, the same topic considered here could be in principle studied in $d>1$. We know~\cite{corsmal2023ordering}, that in $d=2$ the one-time properties of the model are quite different from the $1d$ case, so we expect new scenarios to emerge.
Also, the response function is another interesting two-time quantity, whose properties are important also in connection with the two-time correlations because of the so-called fluctuation-dissipation relation, whereby a possible definition of an effective temperature springs out. This and other issues, like, e.g., the study of the persistence properties \cite{Ben1996},    
remain for future investigations.

\section*{Acknowledgments}

F.C. acknowledges financial support by MUR PRIN 2022 PNRR.

\section*{References}

\bibliography{LibraryStat}

\bibliographystyle{apsrev4-2}

\end{document}